\documentclass[twocolumn,prl,aps,showpacs,nofootinbib,superscriptaddress]{revtex4}
\usepackage{epsfig}
\usepackage{amsmath}

\newcommand{\PR}{{ Phys. Rev.}~}
\newcommand{\PRL}{{ Phys. Rev. Lett.}~}

\begin{document}

\title{
Quantum Dot Version of Berry's Phase: Half-Integer Orbital Angular
Momenta}

\author{V.D. Mur}
\affiliation{Moscow
  Engineering Physics Institute, 115409 Moscow, Russia,}

\author{N.B. Narozhny}
\affiliation{Moscow
 Engineering Physics Institute, 115409 Moscow, Russia,}

\author{A.N. Petrosyan}
\affiliation{Moscow Engineering Physics Institute, 115409 Moscow,
Russia,}

\author{Yu.E. Lozovik}
\affiliation{Institute of Spectroscopy of Russian Academy of
Science,
  Troitsk, 142190 Moscow region, Russia}

\begin{abstract}
We show that Berry's geometrical (topological) phase for circular
quantum dots with an odd number of electrons is equal to $\pi$ and
that eigenvalues of the orbital angular momentum run over
half-integer values. The non-zero value of the Berry's phase is
provided by axial symmetry and two-dimensionality of the system. Its
particular value ($\pi$) is fixed by the Pauli exclusion principle.
Our conclusions agree with the experimental results of T. Schmidt
{\it at el}, \PR B {\bf 51}, 5570 (1995), which can be considered as
the first experimental evidence for the existence of a new
realization of Berry's phase and half-integer values of the orbital
angular momentum in a system of an odd number of electrons in
circular quantum dots.
\end{abstract}

 \maketitle

It is known for a long time already \cite{MSW,Trimer}, see also
\cite{VV}, that in certain cases half-integer quantization of the
orbital angular momentum occurs in molecules. In Ref.~\cite{Trimer}
half-integer values of the orbital angular momentum are associated
with the Berry's geometrical phase $\pi$ that the nuclear wave
function acquires under a pseudo-rotation around the equilateral
configuration of the molecule Na$_3$. This was apparently the first
experimental verification of the Berry's phase in high-resolution
molecular spectroscopy. For reviews on Berry's phase \cite{B,S} in
more general context see Refs.~\cite{SW,Zw}.

In this paper we show that the half-integer quantization of the
orbital angular momentum may occur also in circular quantum dots
with odd numbers of electrons. In these systems the electron motion
may be considered as being restricted to two dimensions. An
assumption of axial symmetry of the confining potential ascertains
the existence of loops that are not deformable to a point: the
topology of the system is equivalent to that of a once-punctured
plane. Hence there exists a topological Berry's phase. Since the
Berry's phase defines a one-parametric set of self-adjoint
generators of rotations, it determines the rotational dynamics of
the electronic system, cf. \cite{Wightman}. The particular value of
the Berry's phase follows from the Pauli exclusion principle. In
circular quantum dots with an odd number of electrons the Berry's
phase takes the value $\pi$ (similarly to the case of the sodium
trimer \cite{Trimer}). Below we demonstrate that our conclusions
agree with experimental results \cite{exp}. Based on our analysis
presented in this paper, we believe that Ref.~\cite{exp} may be
considered as an observation of the new realization of the Berry's
phase.

According to \cite{Havr} (see also \cite{BM}) the oscillatory model
with the parabolic confinement
\begin{equation}\label{pot}
V_{cf}=\sum\limits_{a=1}^Nm_*\Omega^2\mathbf{r}_a^2/2+V^{(0)}\,,
\end{equation}
is a good approximation for low-lying levels in real circular
$N$-electron quantum dots \cite{Ash,Kou}. Here $m_*$ is the
effective mass, $\mathbf{r}_a$ is the two-dimensional coordinate of
an electron, and the effective confining frequency $\Omega$ and the
reference energy level $V^{(0)}$ are phenomenological parameters. In
general $\Omega$ and $V^{(0)}$ depend on the number of particles in
a quantum dot (cf. Ref.~\cite{Havr}) and the quantum numbers
describing the state of the electronic system. Within this approach
the Schr\"{o}dinger equation can be written in dimensionless
variables as
\begin{widetext}
\begin{equation}\label{Eq}
\bigg\{-\frac{1}{4Q^2}\sum\limits_{a=1}^N\frac{\partial^2}
{\partial\boldsymbol{\rho}^2}+\sum\limits_{a=1}^N\rho_a^2+\frac{1}{2}
\sum\limits_{a\neq
b}^N|\boldsymbol{\rho}_a-\boldsymbol{\rho}_b|^{-1}\bigg\}\Psi_N=
\varepsilon(N)\Psi_N.
\end{equation}
\end{widetext}
Here,
\begin{equation}\label{Q}
Q=\frac{1}{\hbar}\bigg(\frac{m_*e^4}{2\epsilon^2\Omega}\bigg)^{1/3}=
\bigg(\frac{\mu}{\epsilon^2}\bigg)^{1/3}\bigg(\frac{E_B}{\hbar\Omega}
\bigg)^{1/3}
\end{equation}
is the dimensionless parameter \cite{QLoz} which is determined by
the ratio of the characteristic Coulomb energy of electron-electron
interaction to the mean level spacing in the confining potential. We
employ the following notations: $\epsilon$ is the effective
dielectric constant, $\mu=m_*/m_e$, $a_B$ is the Bohr radius,
$E_B=m_ee^4/2\hbar^2$, $-e$ and $m_e$ are charge and mass of a bare
electron. The reduced energy $\varepsilon(N)=[E(N)-V^{(0)}(N)]/E_0$
and the dimensionless coordinate $\boldsymbol{\rho}_a
=\mathbf{r}_a/a_0$ are determined by the characteristic energy and
the characteristic size of the system
\begin{equation}\label{char}
E_0=Q\hbar\Omega=\frac{\mu E_B}{\epsilon^2 Q^2}, \,\,
a_0=\left(\frac{2e^2}{\epsilon
m_*\Omega^2}\right)^{\frac{1}{3}}=\frac{2\epsilon}{\mu}\,Q^2a_B.
\end{equation}

Formally, Eq.~(\ref{Eq}) is equivalent to the Schr\"{o}dinger
equation for $N$ particles with the mass $2Q^2$. In the classical
limit $Q\rightarrow\infty$ it determines the equilibrium
configuration of electrons corresponding to the minimum of the
potential energy. For example, in a three-electron quantum dot the
electrons would be located at the vertices of an equilateral
triangle at the distance $\rho_0=(2\sqrt{3})^{-1/3}$ off its center.
The corresponding reduced energy is
$\varepsilon_{cl}(3)=3\sqrt{3}/2\rho_0$. For finite $Q$ corrections
to $\varepsilon_{cl}(3)$ (harmonic and anharmonic) may be calculated
by means of the $1/Q$-expansion \cite{LMN}. For the first three
terms of the $1/Q$-expansion for the energy of the ground state of a
three-electron quantum dot we have \cite{LMN}
\begin{equation}\label{e_3}
\varepsilon_M(3)=3.9311+3.0908Q^{-1}+(0.1908M^2+0.0284)Q^{-2},
\end{equation}
where $M$ is an eigenvalue of generator of rotations
$\hat{L}=-i\partial/\partial\varphi$, $\varphi$ is the angle of
rotation of the system as a whole, $0\leq\varphi< 2\pi$.

It is well known that the differential operator $\hat{L}$ becomes
self-adjoint, i.e. determines an observable, if it is defined on the
Hilbert space of wave functions obeying boundary conditions, which
in their most general form read \cite{Wightman},
\begin{equation}\label{bc}
\Psi(2\pi)=e^{i\theta}\Psi(0), \quad 0\leq \theta< 2\pi\,,
\end{equation}
(see also a recent paper \cite{KPR}). The phase $\theta$ arising as
a result of rotation of the system around the axis of symmetry by
$2\pi$ may be called the Berry's geometrical (topological) phase.
Usually, the Berry's phase \cite {B,S} is acquired by a wave
function in the process of evolution of a system determined by a
Hamiltonian. Here the topological phase $\theta$ itself determines
an operator $\hat{L}_\theta$  from a one-parameter family of
self-adjoint operators and hence the unitary operator
$\hat{U}_\theta$ which describes the rotational dynamics of the
system similarly to the evolution operator,
$$\Psi(\varphi+\tau)=\hat{U}_\theta(\tau)\Psi(\varphi)
=\exp(i\tau \hat{L}_\theta)\Psi(\varphi)\,.$$

In virtue of Eq.~(\ref{bc}), the eigenvalues of the generator
$\hat{L}_\theta$ are given by
\begin{equation}\label{eigv}
 M=\gamma+m, \,\, m =0,\pm 1,\dots,\,
 \theta=2\pi\gamma\,,0\leq\gamma< 1\,.
\end{equation}
The corresponding eigenfunctions
\begin{equation}\label{eigf}
\Psi_M(\varphi)=\exp(iM\varphi)/\sqrt{2\pi}
\end{equation}
implement an irreducible representation of the two-dimensional
rotation group.

Due to axial symmetry, the wave functions in our problem are not
eigenfunctions of the 3D angular momentum operator but of only its
projection (represented by $\hat{L}_{\theta}$) on the axis of
rotation. Hence, according to (\ref{eigv}) $\gamma$ is in principle
an arbitrary number, compare, e.g., \cite{MSW}. Its specific value
is determined by additional physical reasons. If we require that the
wave function remains unaltered after the rotation of the system by
$2\pi$, then $\gamma=0$ and the eigenfunctions (\ref{eigf})
implement a single-valued representation of $O(2)$. In this case the
orbital angular momentum eigenvalues are (up to the factor of
$\hbar$) integers. In cases where $\gamma$ is a rational number the
representation is multiple-valued and the momentum quantization may
be fractional. However, if the system is invariant with respect to
time inversion, only two cases $\gamma=0$ or $\gamma=1/2$ can be
realized \cite{KPR}.

We now go back to the problem (\ref{Eq}) with the odd number of
electrons $N$. Consider first the case $Q\rightarrow\infty$. The
ground state is realized by a rigid configuration of electrons
minimizing the potential energy. This state is invariant under the
$2\pi$ rotation around the symmetry axis. This may be used to
understand the quantization of the angular momentum operator.
Indeed, the overall phase acquired by the ground state wave function
after the rotation is determined by the total momentum $J$. The
$2\pi$ rotation of a two-dimensional system is obviously the
identity element of the symmetric group $S_N$ and thus belongs to
the alternating group $A_N$ of even permutations of the set
$\{1,...,N\}$. Hence, the $2\pi$ rotation of the system is
equivalent to an even number of pairwise transpositions, and
according to the Pauli exclusion principle the wave function do not
change:
\begin{equation}\label{jm1/2}
\exp(i2\pi J)=1\,,\quad J=M+\Sigma=0,\pm 1,\pm 2,\ldots\,.
\end{equation}
Here the total momentum $J$ is represented by the sum of the orbital
and spin angular momenta. Since the number of electrons is odd, the
spin quantum number $\Sigma$ is half-integer. Thus the orbital
angular momentum $M$ must also take a half-integer value. According
to Eq.~(\ref{eigv}) this implies that $\gamma=1/2$ or that the
system is characterized by the Berry's phase $\pi$. To arrive to
this conclusion we have considered the classical limit
$Q\rightarrow\infty$. However, if one varies the parameter $Q$
adiabatically, the quantum numbers $M$ and $\Sigma$ cannot change.
Therefore our result is valid also at $Q\sim1$ which is typical for
real quantum dots.

Consider now the case of three electrons. If the total spin number
$\Sigma=\pm 1/2$, then $M$ can take any half-integer value. However,
the situation is different if $\Sigma=\pm 3/2$. Then there is an
additional symmetry in the problem. The symmetry group of such
system is $C_{3v}$ which is isomorphic to the symmetric group $S_3$.
The group $C_{3v}$ consists of rotations about the symmetry axis by
multiples of the angle $2\pi/3$ (the $C_3$ group) and reflections in
the three bisectrices of the triangle. $C_3$ is isomorphic to $A_3$
and thus the wave function of the system at $Q\rightarrow\infty$
does not change also if it is subjected to a rotation by $2\pi/3$.
Thus,
\begin{equation}\label{jm3/2}
 \exp(iJ2\pi/3)=1\,,\quad J=M+\Sigma=0,\pm 3,\pm 6,\ldots\,.
\end{equation}
This means that $M$ can take the values $M=\pm(3+6k)/2$,
$k=0,1,2,\ldots\,$.

Now we show that our conclusions are in excellent agreement with the
experiment of T. Schmidt {\it et al}. \cite{exp}. The authors of
Ref.~\cite{exp} measured the ground state energy $E(N)$ of
$N$-electron circular quantum dots in GaAs-based heterostructures in
a perpendicular magnetic field $0\leq H\leq16\,T$ for $N=1\div30$.
To explain the data, we should modify the calculation of the
spectrum of a quantum dot in order to take into account the magnetic
field \cite{footnote}.  For this purpose it is enough to make the
changes in Eqs.~(\ref{Eq})-(\ref{char}): (i)
$\Omega(N)\rightarrow\Omega_L(N)$, where
$\Omega_L=\sqrt{\Omega^2+\omega_L^2}$ and $\omega_L=eH/2m_*c$ is
Larmor frequency; (ii) $Q\rightarrow
Q_L=(\mu/\epsilon^2)^{1/3}(E_B/\hbar\Omega_L )^{1/3}$; and (iii)
take into account the Zeeman shift. This way we find for the energy
of a quantum dot
\begin{eqnarray}
\label{E3_H}
&&
E_{M\Sigma}(N;H)=\varepsilon_M(N;H)Q_L\hbar\Omega_L(N)
\\
&&
\nonumber\\
&&
\quad\quad\quad\quad\quad\quad\quad\quad
-(M+\mu g\Sigma)\hbar\omega_L+V_{M\Sigma}^{(0)}(N)\,,
\nonumber
\end{eqnarray}
where $g$ is the effective Lande factor and $\varepsilon_M(N;H)$
replaces $\varepsilon_M(N)$ in Eq.~(\ref{Eq}) after the change
$Q\rightarrow Q_L$. The first three terms of $1/Q$-expansion for
$\varepsilon_M(3;H)$ are given by Eq.~(\ref{e_3}) with $Q$ replaced
by $Q_L$.
\begin{figure}[ht]
\epsfxsize=6cm \centerline{\epsffile{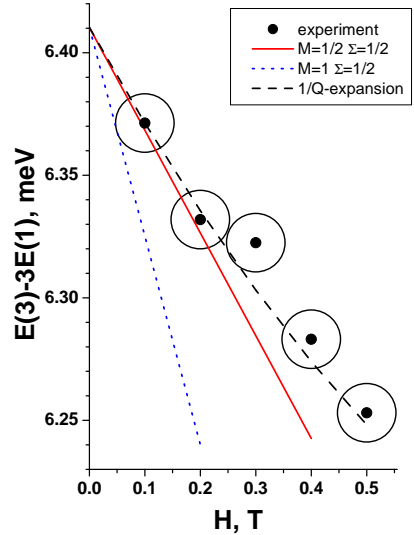}}\caption{The
effective Coulomb energy $E(3)-3E(1)$ versus magnetic field $H$ for
weak field, $\omega_L\ll\Omega(3)$. The dashed line gives the
results of calculations according to Eq.~(\ref{E3_H}) in
approximation (\ref{e_3}) with parameters $\hbar\Omega(3)=4.69\,
meV$, $V^{(0)}(3)=-17.44\,meV$ and $\hbar\Omega(1)=3.60\,meV$. The
experimental points are taken from Ref.~\cite{exp}. The size of the
circles corresponds to the size of experimental points in Fig.~3 of
Ref.~\cite{exp} and reflects experimental error of approximately
$0.025\,meV$. The solid and the dashed lines shows the slope of
$E(3,H)-3E(1,H)$ at $H\rightarrow 0$ calculated according to
Eq.~(\ref{deriv}) for $M=1/2$ and $M=1$ respectively.} \label{weakh}
\end{figure}

It is very important that the first term in the RHS of
Eq.~(\ref{E3_H}) depends on $H^2$. Therefore, in a weak field,
$\omega_L\ll\Omega(3)$, the derivative of energy with respect to
magnetic field is determined for the most part by the Zeeman energy
\begin{equation}\label{deriv}
\frac{dE_{M\Sigma}}{dH}\bigg|_{H=0}=-\frac{e\hbar}{2m_*c}(M+\mu
g\Sigma)\,,
\end{equation}
and does not depend on the shape and parameters of the confining
potential. In the experiment \cite{exp} the typical values of the
parameters are $\epsilon=12.5$, $\mu=0.067$, and $g=0.44$. Thus the
coefficient in the RHS of Eq.~(\ref{deriv}) is equal to $0.864$
(measuring $E_{M\Sigma}$ in $meV$ and $H$ in $T$). On can see from
Eq.~(\ref{E3_H}) that at $H\approx 0.5T$ the quadratic term in the
expansion of $E_{M\Sigma}$ becomes of the order of the Zeeman
energy. Therefore the weak-field interval is $0\leq H<0.5T$.

We calculate the effective Coulomb energy $E(3,H)-3E(1,H)$ which was
measured in the experiment \cite{exp} and show its weak-field
dependence (dashed line) in Fig.~\ref{weakh}. For the energy of a
one-electron quantum dot we adopt the expression
$E(1,H)=\hbar\Omega_L(1)-\mu g\hbar\omega_L/2$, where
$\hbar\Omega_L(1)$ is Fock-Darwin energy and
$\hbar\Omega(1)=3.60\,meV$ \cite{exp}. The experimental points are
taken from Ref.~\cite{exp}. The diameter of the circles corresponds
to the size of experimental points in Fig.~3 of Ref.~\cite{exp} and
reflects experimental error of approximately $0.025\,meV$. Two other
lines in Fig.~\ref{weakh} show the slope of $E(3,H)-3E(1,H)$ at
$H\rightarrow 0$ calculated according to Eq.~(\ref{deriv}). For the
solid line $M=1/2$ and for the dotted one $M=1$. It is clear from
Fig.~\ref{weakh} that, unlike the value $M=1$ for the orbital
angular momentum considered by the authors of Ref.~\cite{exp}, the
value $M=1/2$ agrees with the data quite well. Deviation of
experimental points from the linear law (\ref{deriv}) at $H>0.2\,T$
is explained by influence of the quadratic term in the weak field
expansion of $E_{M\Sigma}$.

The agreement between the data and our calculations leads us to
believe that the results of the experiment \cite{exp} unambiguously
specify the quantum numbers of the ground state of a three-electron
quantum dot right up to the point of the first crossing, or up to
such value of magnetic field $H^{(cr)}$ when the symmetry of the
ground state is changed \cite{Chapl}. Quantum numbers of the ground
state after the first crossing cannot be chosen {\it a priori}
because of the unknown dependence of the phenomenological parameters
$\Omega$ and $V^{(0)}$ in Eq.~(\ref{pot}) on the quantum numbers.
Varying these parameters we can obtain an excellent fit of the
experimental data \cite{exp} by the results of the $1/Q$-expansion.
Certainly, in the experiment \cite{exp} $Q\sim1$, or are even
slightly less than $1$. However, it was shown in Ref.~\cite{LMN}
that for the case of two-electron dots the first three terms of
$1/Q$-expansion provide $3\%$-accuracy even at $Q\lesssim1$. Since
the relative contribution of the Coulomb repulsion for
three-electron quantum dots is greater than for two-electron dots,
we believe that the accuracy of approximation (\ref{e_3}) in the
region $Q\sim1$ is at least of the same order.

\begin{figure}[ht]
\vspace{2.5cm} \epsfxsize10cm \centerline{\epsffile{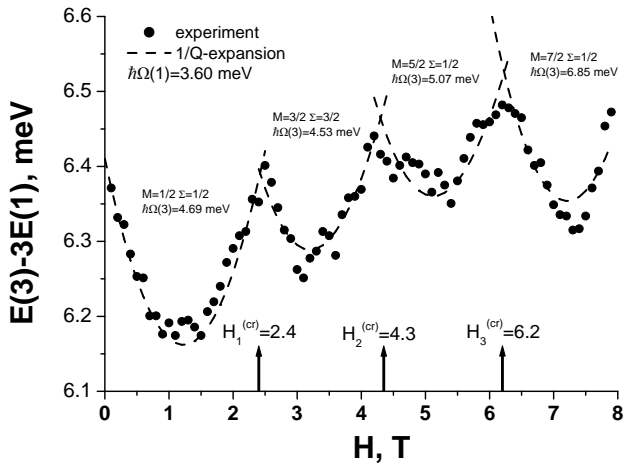}}
\setlength{\abovecaptionskip}{-2.2cm} \caption{\small Comparison of
the experimental data (solid circles) of T. Schmidt {\it at el}.
\cite{exp} for $E(3)-3E(1)$ with the results of calculations
according to Eq.~(\ref{E3_H}) in approximation (\ref{e_3}). The
numbers give the values of the total electron spin and orbital
angular momenta, and the effective confining frequencies. Arrows
indicate the crossing points.} \label{total}
\end{figure}

The result of the fitting procedure is shown in Fig.~\ref{total}.
The experimental points are taken from Ref.~\cite{exp}. We have
found the locations of three crossings in the range $0<H<8T$ at
$H_1^{(cr)}=2.4T, H_2^{(cr)}=4.3T$ and $H_3^{(cr)}=6.2T$. Quantum
numbers $M$ and $\Sigma$ are chosen according to condition
(\ref{jm1/2}) everywhere except the region between the second and
the third crossings, $H_2^{(cr)}<H<H_3^{(cr)}$, where condition
(\ref{jm3/2}) was used. The values of effective confining
frequencies $\hbar\Omega(3)$ are given in Fig.~\ref{total}, and the
values of the parameter $V^{(0)}(3)$ in $meV$ for successive
intervals between the crossing points are $-17.44, -16.23, -20.23,
-35.81$. One can see that the theoretical curve is in a very good
agreement with the experiment.

We believe that the results shown in Figs.~\ref{weakh}, \ref{total}
represent convincing evidence in favor of our interpretation of the
experiment \cite{exp}. Therefore, we regard the data presented in
Ref.~\cite{exp} as the first experimental demonstration of the
existence of theoretically admissible half-integer values of the
orbital angular momentum in two-dimensional quantum systems. We are
also inclined to believe that the appearance of the Berry's phase
$\pi$ and half-integer quantization of the angular momentum in the
experiment \cite{Trimer} arise due to exactly the same physical
reasons as in the case of two-dimensional circular quantum dots.
This is explained by the fact that the configuration of molecule
Na$_3$ (sodium trimer) considered in Ref.~\cite{Trimer} coincides
with the configuration of the system of three electrons in a quantum
dot at $Q\gg 1$ when all electrons are located in the vertices of an
equilateral triangle.

In conclusion, we have predicted the existence of a new version of
Berry's phase along with half-integer quantization of the orbital
angular momentum for 2D axially symmetric systems with an odd number
of confined electrons. We argue that the experimental data for
circular quantum dots in a strong magnetic field \cite{exp} is in
agreement with this statement.

Authors are grateful to V.F. Elesin, A.M. Fedotov, B.N. Narozhny,
V.N. Sobakin, and V.P. Yakovlev for stimulating discussions. This
work was supported by the Russian Foundation for Basic Research
(grant 06-02-17370-a, 07-02-01116-a), by the Ministry of Science and
Education of Russian Federation (grant RNP 2.11.1972), and by the
Russian Federation President grant NSh-320.2006.2.


\begin{thebibliography}{99}
\bibitem{MSW}J. Moody, A. Shapere, and F. Wilczek, \PRL {\bf 56}, 893 (1986).
\bibitem{Trimer}H. von Busch {\it et al}., \PRL {\bf 81}, 4584 (1998).
\bibitem{VV}J. H. Van Vleck, \PR {\bf 33}, 467 (1929).
\bibitem{B}M. Berry, Proc. R. Soc. London A {\bf 392}, 45 (1984).
\bibitem{S}B. Simon, \PRL {\bf 51}, 2167 (1983).
\bibitem{SW}{\it Geometric Phases in Physics}, edited by A. Shapere and
F. Wilczek (World Scientific, Singapore, 1989).
\bibitem{Zw}J. W. Zwanziger, M. Koenig, and A. Pines, Annu. Rev. Phys. Chem.
{\bf 41}, 601 (1990).
\bibitem{Wightman}{A. S. Wightman \it Introduction to Some Aspects of Relativistic
Dynamics of Quantized Fields}, (Princeton Press, Princeton, 1964).
\bibitem{exp}T. Schmidt {\it et al}., \PR B {\bf 51}, 5570 (1995).
\bibitem{Havr}P. Hawrylak, \PRL {\bf 71}, 3347 (1993).
\bibitem{BM}N. A. Bruce, and P. A. Maksym, \PR B {\bf 61}, 4718 (2000).
\bibitem{Ash} R. C. Aschoori, Nature {\bf379}, 413 (1996).
\bibitem{Kou}L. P. Kouwenhoven {\it et al}., Science {\bf 278}, 1788
(1997).
\bibitem{QLoz}A. V. Filinov, M. Bonitz, and Yu. E. Lozovik, \PRL {\bf 86}, 3851 (2001).
\bibitem{LMN}Yu. E. Lozovik, V. D. Mur, and N. B. Narozhny, Zh. Eksp. Teor. Fiz.
{\bf 123}, 1059 (2003) [JETP {\bf 96}, 932 (2003)].
\bibitem{KPR}K. Kowalski, K. Podlaski, and J. Rembieli\'{n}ski, \PR A {\bf 66},
032118 (2002).
\bibitem{footnote}We use the symmetric gauge for vector potential $\,\,\mathbf{A}\,=
\,1/2\,[\mathbf{H}\times\mathbf{r}]$.
\bibitem{Chapl}M. Wagner, U. Merkt, and A. V. Chaplik, \PR B {\bf 45}, 1951 (1992).

\end{thebibliography}
\end{document}